# Agentic AI for Ultra-Modern Networks: Multi-Agent Framework for RAN Autonomy and Assurance

Sukhdeep Singh, Avinash Bhat, Shweta M, Subhash K Singh, Moonki Hong, Madhan Raj K,
Kandeepan Sithamparanathan, Sunder A. Khowaja, Kapal Dev

*Abstract*—The increasing complexity of Beyond 5G and 6G networks necessitates new paradigms for autonomy and assurance. Traditional O-RAN control loops rely heavily on RIC-based orchestration, which centralizes intelligence and exposes the system to risks such as policy conflicts, data drift, and unsafe actions under unforeseen conditions. In this work, we argue that the future of autonomous networks lies in a multi-agentic architecture, where specialized agents collaborate to perform data collection, model training, prediction, policy generation, verification, deployment, and assurance. By replacing tightly-coupled centralized RIC-based workflows with distributed agents, the framework achieves autonomy, resilience, explainability, and system-wide safety. To substantiate this vision, we design and evaluate a traffic steering use case under surge and drift conditions. Results across four KPIs: RRC connected users, IP throughput, PRB utilization, and SINR, demonstrate that a naive predictor-driven deployment improves local KPIs but destabilizes neighbors, whereas the agentic system blocks unsafe policies, preserving global network health. This study highlights multi-agent architectures as a credible foundation for trustworthy AI-driven autonomy in next-generation RANs.

## I. INTRODUCTION

The evolution of Beyond 5G (B5G) and 6G networks is pushing RANs toward unprecedented complexity, with massive device densities, heterogeneous service requirements, and strict performance demands across URLLC, eMBB, and mMTC. To address these challenges, AI/ML frameworks have been increasingly adopted for predictive optimization in traffic steering, handover, interference control, and resource allocation [1] [2]. Architectures such as the O-RAN specification, with Near-RT and Non-RT RICs hosting xApps and rApps, have further advanced programmability and vendor-agnostic intelligence [3] [4].

However, RIC-centric approaches exhibit critical limitations [5] [6]. Decision-making remains centralized, creating bottlenecks and restricting true edge autonomy. AI/ML models deployed within RICs are prone to drift and environmental uncertainties, leading to unsafe or biased policy actions. Moreover, current workflows lack independent verification, allowing policies to be deployed without systematic assurance. Together, these issues raise concerns of scalability, safety, and operator trust in mission-critical deployments, motivating the need for new paradigms of autonomy and assurance.

**Motivation.** These limitations highlight the need for a fundamental shift in how intelligence and assurance are delivered in RANs. Instead of relying on tightly coupled RIC controllers, which centralize decision-making despite their modular design, we argue for a multi-agentic paradigm, where specialized agents collaborate in a distributed manner to collectively achieve autonomy, assurance, and explainability. Each agent is designed with autonomy, interoperability, resilience, and transparency as core principles. For example, agents can independently handle data collection, feature engineering, model training, policy generation, baseline verification, drift detection, secure deployment, and auditing. In this vision, the assurance is not provided by a centralized RIC, but by the cooperative behavior of multiple agents embedded across the network. This approach removes the bottleneck of centralized RIC control, enhances scalability, and builds operator trust by ensuring that only policies validated by independent agents are allowed into deployment.

**Contributions.** This paper makes the following contributions:

- Multi-Agentic Architecture: We propose a distributed multi-agent framework as an alternative to RIC-based control, where agents handle the complete lifecycle of autonomy and assurance in RANs.
- Design of Specialized Agents: We define the internal roles and workflows of agents for orchestration, data collection, feature processing, model training, prediction, policy generation, verification, deployment, auditing, and security.
- Use Case Demonstration: Using a two-cell traffic steering scenario with surge and drift, we show how multi-agentic assurance prevents unsafe policy deployment that would otherwise destabilize neighboring cells.
- Experimental Validation: We present results across four KPIs: RRC connected users, throughput, PRB utilization, and SINR, demonstrating the contrast between naive AI-

S. Singh, A. Bhat, Shweta M, SK Singh, are with Samsung R&D India Bangalore, M. Hong with with Samsung Research, Seoul Korea, K. Sithamparanathan is with The Royal Melbourne Institute of Technology, Australia, S A Khowaja is with Dublin City University, Glasnevin Campus, Kapal Dev is with Munster Technological University, Ireland

- only deployment and agentic verification.
- Autonomy and Assurance Blueprint: We provide a blueprint for future networks, highlighting how multi-agentic architectures can deliver autonomy, safety, resilience, and explainability without reliance on RICs, paving the way for scalable deployment in B5G/6G ecosystems.

## II. Proposed Idea: Agentic AI for Ultra-Modern Networks

### A. System Design

The proposed framework shifts from centralized RIC-based control to a distributed multi-agentic architecture that ensures autonomy, assurance, and explainability across the RAN lifecycle in B5G/6G networks. Instead of relying on tightly coupled controllers, specialized agents independently handle data collection, training, prediction, policy generation, verification, and deployment, while coordinating through standardized protocols for resilience and transparency. Crucially, policies are not deployed immediately after prediction; they are validated against historical baselines or simulated references by independent agents, ensuring safety before execution. By embedding assurance directly into the workflow and distributing intelligence, the architecture removes RIC bottlenecks, enhances scalability, and achieves adaptive closed-loop automation. The design is guided by five principles:

- Autonomy: Agents can execute tasks and recover from failures independently without disrupting the global workflow.
- Interoperability: Standardized inter-agent communication enables seamless coordination across diverse functions.
- Explainability: Every decision is accompanied by a human-readable rationale, strengthening operator trust.
- Resilience: Failures in one agent do not compromise the system; tasks are dynamically rerouted by the Orchestrator.
- Feedback loops: Agents such as the Verifier Agent (VA) and Drift Detector Agent (DDA) continuously refine upstream processes.

Figure 1 illustrates the overall design. The architecture is composed of the following agents, each with specific roles, inputs, outputs, and internal components.

*Detailed Agent Descriptions*

*1. Orchestrator Agent (OA):* Input: Triggers such as anomaly alerts, KPI degradation signals, scheduled optimization events, or operator requests. Output: Workflow instructions for downstream agents specifying which KPIs, cells, or regions to process. Internal Components: Trigger Handler (captures events from monitoring systems or operators), Workflow Manager (maps triggers to task sequences), Conflict Resolver (handles overlapping goals), and Task Dispatcher (delegates and monitors tasks). These components ensure adaptive orchestration and conflict-free workflow execution.

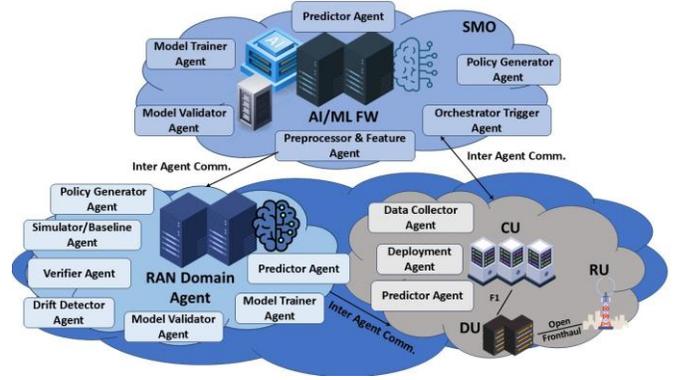

Fig. 1: Holistic View of an idea

*2. Data Collector Agent (DCA):* Input: Raw telemetry streams from CU/DU/RU elements (PRB utilization, SINR, RSRP, throughput, RRC connected users). Output: Structured, timestamped, and schema-compliant telemetry streams enriched with metadata. Internal Components: Connector Module (interfaces via O1, A2, or E2 APIs), Buffering and Windowing (ensures temporal alignment), Schema Validator (normalizes KPI names, units, timestamps), and Integrity Checker (detects and recovers missing or corrupted frames). These ensure synchronized, validated, and reliable telemetry ingestion.

*3. Preprocessor and Feature Agent (PFA):* Input: Structured telemetry from the DCA. Output: Cleaned and feature-engineered datasets for model consumption. Internal Components: Cleaner (removes noise, fills missing values, normalizes data), Feature Extractor (derives moving averages, ratios, temporal gradients), Window Manager (aligns samples into fixed-length time windows), and Anomaly Tagger (flags outliers for drift analysis). This ensures consistent, interpretable, and high-quality data for downstream modeling.

*4. Model Trainer Agent (MTA):* Input: Preprocessed datasets, ground-truth KPI labels, and retraining triggers from the DDA. Output: Optimized, packaged AI/ML models ready for validation. Internal Components: Dataset Manager (splits data into train/validation/test sets), Trainer (executes model optimization), Checkpoint Manager (stores intermediate states for rollback), and Hyperparameter Tuner (balances accuracy and generalization). These enable continuous, robust model evolution [7].

*5. Model Validator Agent (MVA):* Input: Candidate trained models from the MTA and validation datasets. Output: Approved models annotated with performance metrics and robustness indicators. Internal Components: Evaluator (computes accuracy, RMSE, and robustness), Baseline Comparator (assesses improvement over existing models), Safety Filter (detects instability or overfitting), and Approval Engine (flags validated models for deployment). This ensures only reliable models enter operational workflows.

*6. Predictor Agent (PA):* Input: Validated models from the MVA and real-time features from the PFA. Output: Predicted KPI trajectories and scenario-specific forecasts. Internal Components: Model Loader (initializes the latest validated model), Forecaster (generates short- and long-term predictions), Uncertainty Estimator (adds confidence intervals), and Scenario Generator (simulates alternate load/interference conditions). Together, these support proactive network control decisions.

*7. Policy Generator Agent (PGA):* Input: Predicted KPI trajectories from the PA. Output: Candidate network policies (e.g., user offloading, scheduling adjustments). Internal Components: Policy Constructor (encodes actions), Impact Annotator (tags expected KPI impacts), Trade-off Analyzer (highlights performance-resource trade-offs), and Policy Queue (stores policies awaiting verification). This enables structured and explainable policy generation.

*8. Simulator/Baseline Agent (SBA):* Input: Historical KPI datasets and simulation parameters. Output: Reference KPI trajectories representing safe or expected network behavior. Internal Components: Baseline Generator (reconstructs normal KPI trends), Synthetic Simulator (approximates future scenarios), and Comparison Interface (packages results for the Verifier Agent). This maintains reliable baseline projections for policy evaluation.

*9. Verifier Agent (VA):* Input: Candidate policies from the PGA, baseline trajectories from the SBA, and predicted KPIs from the PA. Output: Policy decisions labeled as approved, rejected, or requiring retraining. Internal Components: Comparator (evaluates deviations between predictions and baselines), Threshold Checker (enforces KPI constraints, e.g., PRB utilization or SINR limits), Decision Engine (classifies policy safety), and Feedback Notifier (alerts the DDA of unsafe or anomalous outcomes). The VA ensures that only safe policies proceed to deployment.

*10. Drift Detector Agent (DDA):* Input: Live KPIs, prediction errors, and verifier feedback. Output: Drift alerts and retraining requests to the MTA. Internal Components: Error Monitor (tracks prediction-vs-actual deviation), Drift Test (applies statistical checks), Severity Classifier (categorizes drift levels), and Retraining Trigger (initiates model updates). These maintain long-term system adaptability and stability.

*11. Deployment Agent (DA):* Input: Verified policies from the VA. Output: Enforced configurations on CU/DU/RU nodes. Internal Components: Policy Translator (converts abstract actions into device-specific commands), Executor (applies configurations), Rollback Manager (restores prior states on degradation), and Execution Logger (records timestamps and actions). This ensures safe, auditable, and efficient policy deployment.

*12. Audit and Explainability Agent (AEA):* Input: Logs from the VA, DA, and OA. Output: Operator-facing audit reports and explainable insights. Internal Components: Decision Logger (captures decision context), Explanation Generator (translates reasoning into human-readable format), Compliance Checker (verifies regulatory adherence), and Report Engine (produces dashboards and audit trails [8]). This reinforces transparency and accountability.

*13. Security Agent (SA):* Input: Inter-agent communication streams, policy commands, and operational logs. Output: Verified, tamper-proof communications and integrity alerts. Internal Components: Auth Manager (authenticates and authorizes agents), Integrity Checker (verifies unaltered packet delivery), Threat Monitor (detects anomalies or adversarial behavior), and Recovery Module (isolates compromised agents and informs the OA). This safeguards the reliability and trustworthiness of all agentic operations.

## III. USE CASE OF THE MULTI-AGENT ARCHITECTURE - AI/ML FW POLICY VERIFICATION

One of the most critical challenges in current AI/ML-driven RAN control is that policies predicted by trained models are often deployed directly into the network without robust verification [9]. While such policies may optimize local KPIs in the short term, they can destabilize neighboring cells, cause unintended interference, or degrade end-to-end QoS under conditions of drift or sudden demand surges. To address this, the proposed multi-agent architecture introduces an independent verification layer that assures safety before policy deployment. The proposed multi-agent architecture operates through the following sequence of steps and depicted in Fig. 2.

The workflow begins when the Orchestrator Agent (OA) receives a trigger, such as KPI degradation, anomalies, or scheduled optimization runs. The Data Collector Agent (DCA) gathers raw telemetry (e.g., PRB utilization, SINR, RRC connected users, throughput) from the target and neighboring cells, which the Preprocessor and Feature Agent (PFA) cleans and converts into structured feature sets. These feature sets, along with ground truth labels, are used by the Model Trainer Agent (MTA) to build or refresh AI/ML models. The Model Validator Agent (MVA) evaluates these models, approving only those that are robust and reliable.

Once validated, the Predictor Agent (PA) uses these models to forecast KPI trajectories under candidate optimization actions, such as traffic steering or load balancing. The Policy Generator Agent (PGA) then formulates policies (e.g., offloading percentages, scheduling changes) annotated with their predicted impacts. Instead of deploying these directly, the Simulator and Baseline Agent (SBA) independently generates reference KPI trajectories based on historical data and lightweight simulations.

The Verifier Agent (VA) performs the critical step of comparing the predicted KPI outcomes of the AI/ML framework against these independently generated baselines. If a candidate policy is found to violate safety thresholds (e.g., pushing neighbor PRB utilization above 70% or degrading SINR below acceptable levels), it is rejected, and a feedback loop is initiated. The Drift Detector Agent (DDA) then examines whether the rejection is due to model drift or unseen data distributions and, if necessary, triggers retraining via the MTA. Only policies that pass verification are handed over to the Deployment Agent (DA) for execution in the live network. Finally, the Audit and Explainability Agent (AEA) documents the rationale behind the approval or rejection, while the Security Agent (SA) ensures integrity and trust in inter-agent communication throughout the workflow.

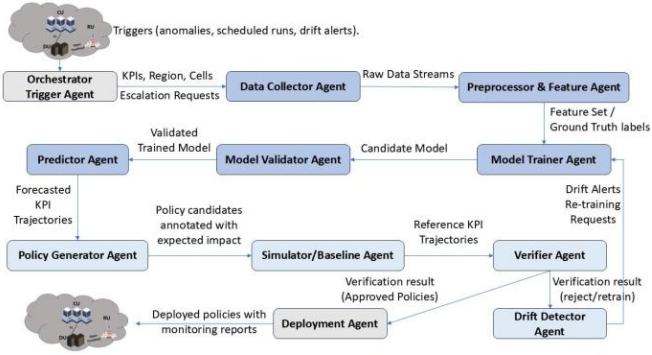

Fig. 2: End-to-End Flow for a Use Case

Through this pipeline, the architecture ensures that policies generated by the AI/ML framework are not blindly deployed. Instead, they are subjected to multi-agent verification and assurance stages that protect the network against unsafe decisions. This use case highlights the strength of agentic collaboration: the AI/ML framework remains a valuable tool for prediction, but its outputs are safeguarded by independent verification agents, guaranteeing that only safe and explainable policies reach deployment.

For clarity, in a non-agent based pipeline, the AI/ML framework directly predicts KPI trajectories and the resulting policies are applied to the network without an independent validation stage. Such a pipeline is vulnerable to unsafe policy deployment when model drift or anomalous conditions occur. In contrast, the proposed multi-agentic pipeline inserts verification and assurance stages (implemented as independent agents) between prediction and deployment, thereby preventing unsafe actions and enabling explicit retraining and human-in-the-loop interventions when required.

## IV. Results and Discussion

### A. Experimental Setup

The experimental validation was conducted using ns-3 and OpenAirInterface (OAI) based simulation testbeds. These environments model end-to-end RAN behaviors including user mobility, radio propagation, interference, scheduling, and control signaling. They enable measurement of KPIs such as throughput, SINR, PRB utilization, and RRC connection counts under realistic scenarios, allowing direct comparison between the baseline (non-agent) pipeline and the proposed multi-agentic pipeline.

To substantiate the proposed architecture, we evaluate it through a traffic steering use case in a two-cell RAN topology. Traffic steering is a common optimization strategy where users are offloaded from a congested target cell to a neighboring cell to balance load [10]. While effective in reducing congestion locally, naive AI-driven deployment of such policies can unintentionally overload neighboring cells, particularly under conditions of environmental drift. This makes traffic steering an ideal use case to demonstrate how the proposed multi-agentic framework ensures assurance by verifying policies before deployment, thereby preventing unsafe actions and preserving overall network stability. The objective of this experiment is to demonstrate how an AI-based traffic steering policy behaves under two contrasting deployment strategies: (a) a naïve approach where the policy is applied directly from a strong predictor, and (b) an *agentic approach* where the policy is subjected to an independent verification stage before deployment, particularly under conditions of environmental drift. The experimental topology consists of two adjacent cells: Target Cell A, which becomes congested due to a user surge, and Neighbor Cell B, which receives offloaded traffic. A total of 200 user equipments (UEs) are distributed initially between the two cells, with 120 attached to Cell A and 80 attached to Cell B. The UEs remain stationary within a trial to isolate load dynamics. A surge scenario is introduced between time $t = 100$ and $t = 140$, during which up to 90 additional users are gradually admitted into Cell A. The simulation runs for 240 minutes in total, with both cells configured with nominal capacities of 200 Mbps (Target) and 150 Mbps (Neighbor).

Per-minute telemetry is collected for four key performance indicators (KPIs): RRC connected users, IP throughput, downlink PRB utilization, and SINR. The KPI models in the simulator explicitly define how load maps to KPI values: PRB utilization increases proportionally with excess RRC users, SINR decreases with rising PRB utilization to capture interference effects, and throughput is approximated as the product of capacity, residual PRB headroom, and an efficiency factor. These deterministic mappings ensure the experiment is reproducible.

In this paper, an agent is defined as a self-contained, context-aware functional entity that performs specific tasks such as data collection, preprocessing, prediction, policy generation, verification, or deployment. Each agent operates based on its role in the control loop, receiving inputs from raw network telemetry (e.g., PRB utilization, SINR, throughput), outputs of other agents, or simulated/historical baselines, and producing standardized policy or KPI messages consumable by peer agents. Internally, agents use lightweight AI/ML models, decision logic, and secure communication protocols to perform their functions. For implementation, the Predictor Agent employs multilayer perceptron (MLP) and LSTM models for temporal KPI forecasting, while the Drift Detector Agent applies Kolmogorov–Smirnov (KS) and CUSUM tests for drift detection. All agents are deployed as independent Python microservices communicating via gRPC-based A2A channels secured with TLS, exchanging structured JSON messages for KPI metrics, policy objects, and verification outcomes. Optimization occurs through local hyperparameter tuning, with system-level adaptation achieved via MARL-inspired feedback between Verifier and Drift Detector agents to enhance policy safety and stability.

The AI predictor was designed as a sequence model, such as an LSTM; however, in the practical experiment a multilayer perceptron (MLP) trained on flattened 10-minute windows was employed as a proxy for a stronger predictor. This model rolled forward autoregressive forecasts over a 10-minute horizon for all four KPIs. The agentic architecture was then layered on top of this predictor. The **Data Collector Agent (DCA)** acquired per-minute telemetry, while the **Pre-processing Agent (PFA)** constructed rolling feature windows. The **Model Training Agent (MTA)** and **Model Validation Agent (MVA)** trained and validated the predictor on pre-surge data. The **Predictor Agent (PA)** generated forecasts, which the **Policy Generator Agent (PGA)** translated into offload fractions (0–50%) whenever predicted PRB utilization exceeded 0.80 or RRC counts exceeded baseline thresholds. Crucially, the **Simulator/Baseline Agent (SBA)** independently simulated the effects of candidate policies using current telemetry, including drifted states. The **Verifier Agent (VA)** then compared the predictor's post-policy forecast with the SBA simulation, enforcing neighbor safety constraints (e.g., simulated neighbor PRB below 0.85). If the policy passed verification, the **Deployment Agent (DA)** applied it to the simulator; otherwise, it was blocked. Finally, the **Audit Agent (AEA)** logged artifacts and outputs.

To introduce risk, a **drift scenario** was injected at $t = 138$, just prior to likely policy deployment. Neighbor Cell B was degraded by increasing its PRB baseline by +0.15, reducing capacity to 60% of original, and decreasing SINR by approximately 2 dB. This represents a sudden degradation, such as maintenance or partial outage, that was not present in the training data. Two scenarios were then executed: (i) **No-Agent**, in which the predictor's policy was deployed directly, and (ii) **Agentic**, in which the SBA and VA determined whether deployment should proceed. All runs produced reproducible artifacts in the form of CSV telemetry and comparison graphs. A 30-minute evaluation window following deployment ($t = 140$ to $t = 169$) was used to compare pre- and post-deployment performance in both scenarios.

The experimental results highlight the contrast between aggressive improvements at the target cell and system-level risks at the neighbor, and how the agentic approach mitigates those risks. Each graph includes three traces: the pre-deployment baseline, the No-Agent deployment outcome, and the Agentic outcome. In the particular run presented, the VA rejected deployment after detecting unsafe consequences for the neighbor, leaving the Agentic trace at the drifted baseline.

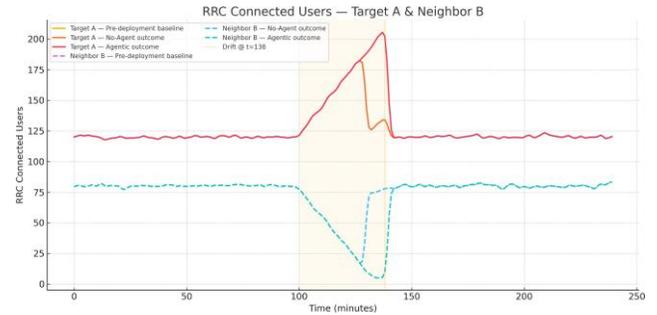

Fig. 3: RRC dynamics under surge and drift

**RRC Connected Users:** In the No-Agent scenario, the target RRC count decreased significantly as excess users were offloaded to the neighbour, with the average falling from approximately 175.1 to 157.1 (10.3%). The neighbor RRC count rose sharply in parallel. In contrast, the Agentic scenario kept both cells at pre-deployment levels since the VA blocked unsafe offloading as shown in Fig. 3. This demonstrates that while No-Agent superficially improves the target cell, it ignores neighbor strain, whereas the agentic pipeline enforces system-wide safety.

**IP Throughput:** Target throughput increased markedly under No-Agent, rising by approximately 20.9% (59.9 Mbps to 72.4 Mbps). However, this came at the cost of a collapse in neighbor throughput due to reduced capacity and overload as shown in Fig. 4. The Agentic system, by refusing deployment, maintained throughput at the drifted baseline, thereby protecting global QoS even at the expense of local gains.

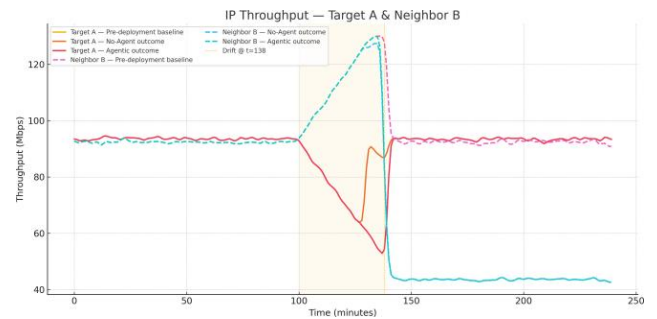

Fig. 4: Throughput impact of offload and drift.

**PRB Utilization:** Resource occupancy reflected the same trade-off. In No-Agent, target PRB utilization dropped from 0.671 to 0.598 (10.9%), but the neighbor's PRB surged from 0.413 to 0.586, with the 95th percentile reaching 0.663 as shown in Fig. 5. The Verifier Agent prevented this unsafe redistribution of load, preserving resource stability at the neighbor.

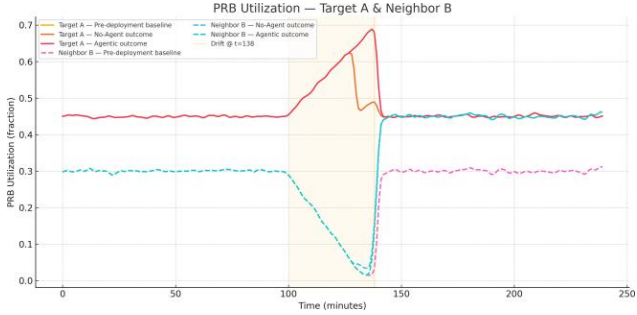

Fig. 5: PRB utilization trends under surge.

**SINR:** Physical layer impacts were also evident. Target SINR improved modestly under No-Agent (+1.14 dB), while the neighbor's SINR degraded severely due to overload as shown in Fig. 6. Agentic deployment avoided additional harm by leaving SINR unchanged at the drifted baseline.

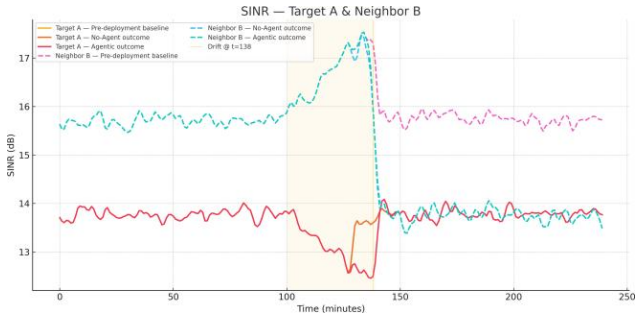

Fig. 6: SINR variations with drift and policy actions.

Across all KPIs, the trade-off is clear. The No-Agent approach produced tangible short-term benefits for the target cell, but these came with severe penalties for the neighbor. The Agentic pipeline, leveraging SBA simulation and VA checks, prevented unsafe deployment and prioritized network-wide stability.

**Conclusion of Experiment:** The findings underscore that even strong predictors such as MLPs or LSTMs are vulnerable to environmental drift. Blindly deploying their recommendations risks improving one cell at the cost of destabilizing others. The agentic architecture, by enforcing verification against current telemetry and neighbor guardrails, prevents such unsafe actions. In this run, Agentic chose to block deployment, sacrificing immediate local gains to protect global network health. The broader implication is that predictive AI in RAN optimization must always be paired with independent verification mechanisms to ensure safety, robustness, and trustworthiness in operational networks.

## V. CONCLUSION

This paper presented a shift from RIC-centric architectures toward a *multi-agentic approach for RAN autonomy and assurance*. We demonstrated that while AI predictors can optimize local KPIs, they may produce unsafe consequences under drift, destabilizing neighboring cells. By contrast, the multi-agent system, through baseline simulation and verification, blocked unsafe policies and preserved global stability. The conclusion is clear: autonomy in B5G/6G cannot rely solely on tightly-coupled AI predictors or centralized RICs; it requires distributed agents working collaboratively to ensure safety, resilience, and explainability.